\documentstyle[prl,aps]{revtex}

\begin{document}
\draft

\twocolumn[
\hsize\textwidth\columnwidth\hsize\csname@twocolumnfalse\endcsname

\title{Hole doping and weak quenched 
disorder effects on the 1D Kondo lattice, for
ferromagnetic Kondo couplings}               
\author{Karyn Le Hur}
\address{Laboratoire de Physique des Solides, Universit{\'e} Paris--Sud,
		    B{\^a}t. 510, 91405 Orsay, France}
 \maketitle

\begin{abstract}
We investigate the one-dimensional Kondo lattice model (1D KLM) for ferromagnetic Kondo 
couplings, by using the bosonization method. The ferromagnetic 2-leg spin ladder and the S=1 
antiferromagnet occur as new one-dimensional 
Kondo insulators, for a 
half-filled band. First, 
a very small hole-doping makes the charge sector massless
and it can produce, either an incommensurate RKKY interaction or
a S=1 ferromagnet according to the strength of 
the ferromagnetic Kondo coupling. Second, we investigate, the effects of a 
weak and quenched disorder on these two Kondo insulators, by applying 
renormalization group methods: the Anderson localization is suppressed
 only in the context of a strong ferromagnetic Kondo coupling. 

\end{abstract}
\vfill
 \twocolumn
\vskip.5pc ]
\narrowtext

The Kondo lattice model (KLM) consists
of conduction electrons magnetically coupled to a spin array via the Kondo 
interaction. This
 model is ruled by two important parameters, namely the hopping strength $t$ of the
 conduction electrons and the Kondo coupling $\lambda_k$, which can be ferromagnetic
 or antiferromagnetic. Many various
 phases are expected in the phase diagram, according to the ratio 
 $\lambda_k/t$. For instance, the compounds which yield a large and positive 
 ratio $\lambda_k/t$ are called
 the ``ancestral'' Kondo insulators, and they behave, at low-temperatures, like semi-conductors 
 with very small gaps\cite{trois}. Conversely, the
 compounds which yield a small and positive ratio $\lambda_k/t$ have generally a magnetic ground
 state. New interesting fixed points
 of the KLM, have also been investigated, for some years ago, like the heavy-fermion one, which corresponds
 to an intermediate (positive) ratio $\lambda_k/t$\cite{quatre}: these heavy-fermion
 compounds are Fermi liquid systems on the brink of magnetism. Exotically, the low-dimensional systems such as the S=1 antiferromagnet\cite{ren,Ww}
 or the 2-leg spin-ladder\cite{no} have also attracted both theorists and experimentalists: curiously, these 
 low-dimensional spin systems yield a similar ground state and for instance, they are known to develop a spin gap in their 
 singlet-triplet excitations.
 
Finally, in this Letter, we investigate the case of the 
1-dimensional KLM (1D KLM) but originally for a {\it ferromagnetic} Kondo coupling. In this context, the ferromagnetic 2-leg spin 
 ladder and the S=1 antiferromagnet occur as new interesting Kondo 
 insulators. Explicitly, in this paper, by applying standard bosonization
 methods\cite{huit,huitb}, we begin to yield clearly these original phases of 
 the 1D KLM, for a half-filled band. Then, we
 view the effects, on these two "similar" Kondo insulators, of a small hole-doping 
 and a weak and dense randomness, by using the well-known replica trick and renormalization-group
 analysis. 
 
 We, first, consider the pure 1d KLM:
 \begin{eqnarray}
 \label{zero}
 {\cal H}&=&-t\sum_{i,\sigma} c^{\dag}_{i,\sigma}c_{i+1,\sigma}+U\sum_{i,\sigma}n_{i,\sigma}n_{i,-\sigma}\nonumber\\
 &+&\lambda_k\sum_i
           \vec{S}_{c,i}.\vec{S}_{f,i}+\lambda_f\sum_i\vec{S}_{f,i}.\vec{S}_{f,i+1}
 \end{eqnarray}          
  with:        $(\lambda_k<0\ ,\ \lambda_f>0\ ,\ U>0)$.
 \\ 
 Here, $c^{\dag}_{i,\sigma}$ $(c_{i,\sigma})$ is the creation (annihilation)
 operator of conduction electrons,  $\vec{S}_c=c^{\dag}_{\alpha}\frac{{\vec\sigma}_{\alpha\beta}}{2}
 c_{\beta}$ with the Pauli matrix ${\bf \sigma}$ and similarly, we introduce $\vec{S}_f=f^{\dag}_{\alpha}\frac{{\vec\sigma}_{\alpha\beta}}{2}
 f_{\beta}$. The term $\lambda_k$ describes the ferromagnetic Kondo 
 coupling. An {\it ad hoc} coupling $\lambda_f$ between the 
 localized spins is introduced and it will be fixed by physical arguments. A Hubbard interaction between electrons has also been
 introduced. We approximate that the lattice step $a\rightarrow 0$ and we use the following conventions on
 the Minkowskian space. The relativistic fermions
 $c_{\sigma}(x)$ are separated in left-movers,  $c_{L\sigma}(x)$ and
 right-movers,  $c_{R\sigma}(x)$ on the $v$-cone. By using the bosonization method, the
 charge $J_{c,L}= c^{\dag}_{L\sigma}c_{L\sigma}(x)$ and spin current  
  $\vec{J}_{c,L}=c^{\dag}_{L\alpha}\frac{{\vec\sigma}_{\alpha\beta}}{2}c_{L\beta}$
operators  can be respectively depicted by a massless
 scalar field $\Phi_c^c$ and a SU(2) matrix g\cite{neuf}; we
 just remember the identifications: $J_{c,R}+J_{c,L}=\frac{1}{\sqrt\pi}\partial_x
 \Phi_c^c$ and $J_{c,R}-J_{c,L}=-\frac{1}{\sqrt\pi}\Pi_c^c$, where $\Pi_c^c$ is the moment conjugate
 to the field $\Phi_c^c$. Then, we decompose the spin operator, into its k=0 and $k=\pi$ components as
 follow\cite{neuf}: 
  \begin{eqnarray}
   \vec{S}_c&\simeq&\vec{J}_{c,L}(x)+\vec{J}_{c,R}(x)\\ \nonumber
   &+&cst\exp(i2k_F x)tr(g.\vec{\sigma})\cos(\sqrt{2\pi}\Phi_c^c)
   \end{eqnarray}
  Concerning the localized spins, $\exp(i\sqrt{2\pi}\Phi_f^c)$ is simply replaced
  by its expectation value; finally, one obtains the general Hamiltonian:
  \begin{equation}
 {\cal H}=H_c+H_s+H_k
 \end{equation} 
 with:
 \begin{eqnarray}
 H_c&=&\int dx\  \frac{u_{\rho}}{2 K_{\rho}}:{(\partial_x\Phi_c^c)}^2:+\frac{u_{\rho}K_{\rho}}{2} :{(\Pi_c^c)}^2:\\ \nonumber
 &+&cst.g_3\cos(\sqrt{8\pi}\Phi_c^c)\\ \nonumber
 H_s&=&\frac{2\pi v}{3}\int\ dx\ :\vec{J}_{c,L}(x)\vec{J}_{c,L}(x):\\ \nonumber 
 &+&\frac{2\pi v_f}{3}\int\ dx\ :\vec{J}_{f,L}(x)\vec{J}_{f,L}(x):+  L\rightarrow R \\ \nonumber
 H_k&=&\int dx\ \lambda_1 \vec{J}_{c,L}(x)\vec{J}_{f,L}(x)
 + \lambda_2 \vec{J}_{c,L}(x)\vec{J}_{f,R}(x)\\ \nonumber
    &+&L\leftrightarrow R + \lambda_3 tr(g^{\dag}.\vec{{\sigma}}) tr(f.\vec{{\sigma}})\cos(\sqrt{2\pi}\Phi_c^c)    
 \end{eqnarray}
 The coupling $g_3\simeq U/t$ generates the usual $4 k_F$-Umklapp process and 
  the $u_{\rho}$, $K_{\rho}$ parameters of the Tomonaga-Luttinger (TL) liquid\cite{neufb}, are given by:
  \begin{equation}
  \label{n}
  u_{\rho}K_{\rho}=v \qquad \text{and} \qquad \frac {u_{\rho}}{K_{\rho}}=v+U/\pi
  \end{equation}
 The {\it dimensionless} parameters $\lambda_{i=1,2,3}$ vary like $(a\lambda_k)$ and the two velocities
 $v_f$ and $v$ are respectively of order $(a.\lambda_f)$ and $(a.t)$. Due to the SU(2) spin symmetry of electrons, we
 remember that the spinon velocity is simply equal to $v$. 
 \\
 The terms $\lambda_1$ and $\lambda_2$
 are ${\it marginal}$: the term $\lambda_1$ just
 renormalizes the velocity of spinons; it can be ignored. Conversely, the term $\lambda_2$ generates
 the ferromagnetic Kondo process and the term $\lambda_3$ describes the coupling
 between the staggered magnetizations. 
    
    When $a\left|\lambda_k\right|<<v$ ($t$ is large), the term $\lambda_2$ 
    is small and it is not really essential at half-filling: it can be forgotten. Conversely, the
    term $\lambda_3$ polarizes the $S_f$-spins and then, we can take $v_f\simeq v$. The $2k_F$ oscillation
    becomes commensurate with the alternating localized 
    spin operator. The term $\lambda_3$ has the dimension 3/2 (in the limit $U<<\lambda_k$)
    and it opens a mass gap $M\propto \lambda_3^2$, for the charge and all the spin 
    excitations. First, the Umklapp process does not change the physics and it can be 
    dropped out. Second, all the spin excitations are gapped because the spin Hamiltonian is not {\it conformally invariant}, when we couple
    2 $SU(2)_{k=1}$ algebras. To briefly prove that, we note $K=\lambda_3 \langle\cos(\sqrt{2\pi}\Phi_c^c)\rangle $ which 
    is proportional to $-(a\lambda_k)^2$; then, by using the representation of 
   $\vec{J}_{f,(L,R)}$ in terms of the matrix f\cite{neuf}, we obtain the {\it standard} 2-leg
   spin-ladder Hamiltonian\cite{cinqs}: 
   \begin{equation}
   H^*=H_s+H_k=W(f)+W(g)+K tr(g^{\dag}.\vec{{\sigma}}) tr(f.\vec{{\sigma}})
   \end{equation}
   with the well-known $SU(2)_{k=1}$ Wess-Zumino-Witten Hamiltonian\cite{dix,onze}:
   \begin{eqnarray}
   W(f)&=&\frac{1}{4 u^2}\int d^2 x\ tr(\partial_{\mu}f \partial_{\mu}f^{\dag})\\ \nonumber
          &+&\frac{k}{24\pi}\int d^3 x\ \epsilon^{\mu\nu\lambda} tr(f^{\dag}\partial_{\mu}f 
           f^{\dag}\partial_{\nu}ff^{\dag}\partial_{\lambda}f)
   \end{eqnarray}
 \\
 We take $v=1$ in the definition of $W(f)$ and the central charge k=1 is not subject to 
 renormalization. Now, by substituting $g^{\dag}=\alpha f^{\dag}$, we find
 that the
 $\alpha$-field action concerns the high energy physics and that
 the asymptotic behavior of the system 
 is then well-described by the action 2W(f)\cite{cinqs,douze}: this effective action
  is not conformally invariant because it 
  is described by a k=0 central charge. Consequently, the u-coupling becomes asymptotically free in the
 Infra-Red limit and it obeys:
 \begin{equation}
  \beta(u)=\frac{d u^2}{d lnD}=-\frac{1}{4\pi}u^4
   \end{equation}
  By integrating this equation, we find a ground state which looks like a {\it short-range} Resonating Valence Bond (RVB) system\cite{AN}
 confined inside the characteristic length 
 $\xi\propto\frac{u^2}{D}\exp(\frac{4\pi}{u^2})$ and D is an ultraviolet cut-off. Another consequence is that
 spin-spin correlation functions
 decay exponentially or equivalently, that the magnetism is generated by the
 massive singlet-triplet excitations: the corresponding mass or more commonly, the Haldane gap
 $\Delta_S\propto \left|K\right|/2$ is well-described by the instantonic term 
 $tr g^{\dag}.tr f\simeq (tr f)^2$\cite{douze} and we can consider that $D\propto \lambda_3^2$. 
 
  When $a\left|\lambda_k\right|>v$ ($t$ is small), the electrons are trapped
  by the lattice spins and the spin system looks like a S=1 antiferromagnet. For a strong ferromagnetic Kondo coupling, the 
  SU(2) spin symmetry of electrons is explicitly broken onto $U(1)\times
 {\cal Z}_2$ and the ${\cal Z}_2$ symmetry which corresponds to $\sigma_z\rightarrow -\sigma_z$ is spontaneously broken. The term $\lambda_2$ is then decomposed
 into $(\lambda_{2,+-},\lambda_{2,z})$ and the Kosterlitz-Thouless analysis 
 predicts that a mass turns on exponentially with anisotropy:
  \begin{equation}
  \label{uu}
    m\propto\lambda_3^2\exp(-cst/\sqrt{\left|{\Delta}J_z\right|})
    \end{equation}
    with: 
   \begin{equation} 
   \lambda_{2,+-}-\lambda_{2,z}=\left|{\Delta}J_z\right|<<0
    \end{equation}
  Remarkably, in that case, it is well-known\cite{ren} that the {\it same} 
  original RVB ground state occurs and finally, the preceding ``weak $\lambda_k$-coupling'' 
  approach still gives good qualitative results 
  in that case: the S=1 antiferromagnet is also ruled by a central charge k=0 and also by
  the term $\lambda_3$. Finally, the mass $m$ of eq. (\ref{uu}) just gives some little details about
  the singlet-triplet excitations (now the massive triplet excitations 
  are reduced to $S^{z}=\pm 1$). In conclusion, by
  decreasing the hopping strength of electrons, we do not hope
   many drastic changes for
  a half-filled band, when the Kondo coupling is ferromagnetic. 
 
 Even for a {\it small} hole-doping, the oscillant term $\lambda_3$ can be dropped 
 out; then, the system behaves like a metallic state and it is
 not well-described by the preceding  short-range RVB ground state, whatever
 the ratio $\left|\lambda_k\right|/t$. In fact, the Haldane
 gap is destroyed by the commensurate-incommensurate transition and we
 can conclude that, conversely to the standard coupled
 spin-chain problem, the spin sector is not well-described by {\it any} RVB ground state away from
 half-filling. Then, according to the ratio $\left|\lambda_k\right|/t$, we obtain
 the following ground states.
\\
 In the regime $a\left|\lambda_k\right|<<v$, the
 {\it ad hoc} $\lambda_f$-coupling is not well-suited to the situation and we start
  with $v_f=0$. The spin system is simply described by a $SU(2)_{k=1}$ algebra 
 but now, we can not omit the small Kondo coupling although it is not
  subject to renormalization ($\lambda_2^*\simeq\lambda_2<<v$). Then, the $SU(2)_{k=1}$ algebra is explicitly described
  by the spin field: $\vec{J'}_L\simeq \alpha\vec{J}_{c,L}+\beta\vec{J}_{f,R}$ and
  this formula is correct while $\beta<<\alpha$. The parameters, $\alpha$ and $\beta$ are given by
  the two following constraints:
  \begin{equation}
  \beta^2+\alpha^2=k=1 
  \end{equation}
  and:
  \begin{equation} 
  \beta/\alpha=3\lambda_2^*/4\pi v
  \end{equation}
  Then, by solving these equations we find that only a very small ratio of 
  the localized spins, $\beta^2=(1+\frac{16{\pi^2 v^2}}{9\lambda_2^{*2}})^{-1}<<1$ 
  becomes itinerant. Then, the Kondo coupling just renormalizes the velocity of 
  spinon to $v^*=v/{\alpha^2}=v.(1+\frac{9\lambda_2^{*2}}{16{\pi^2 v^2}})$. Finally,
  the $S_f$-spins remain localized and the electrons just 
  polarize them by hopping. By using general results on the
  TL liquid, we can predict power-law for the spin-spin correlation functions:
  \begin{equation} 
 \langle \vec{J_f}(0,0).\vec{J_f}(x,0)\rangle\sim\langle \vec{J_c}(0,0).\vec{J_c}(x,0)\rangle\propto
 \frac{\cos(2k_F x)}{x^{1+K_{\rho}}}
 \end{equation}
 The system remains 
 antiferromagnetic but we predict the disappearance of the Haldane 
 gap and the enhancement of the RKKY interaction. 
 \\
  When $a\left|\lambda_k\right|>v$, the
  electrons are trapped by spins and the
 spin system is now composed prevalently of S=1 spins and of some rare spinons. For a {\it very} small hole-doping, 
 a $SU(2)_{k=2}$ algebra is hoped to still describe the ground state and we can not
 use a ``weak $\lambda_k$-coupling'' approach in that case. Anyway, the ground
 state is quite easy to intute. Indeed, since $\lambda_3=0$, we do not
 expect any $k=\pi$ fermionic excitations and finally, this implies that the S=1 spins 
 form a ferromagnetic ground
 state: this is also in agreement with a metallic ground state. 
 \\
 Finally, we
 expect a cross-over from an antiferromagnetic and incommensurate RKKY situation towards 
 a ferromagnetic ground state (formed by the S=1 spins) by decreasing the hopping
 strength $t$ of electrons.
 
 Now, for a half-filled band, we investigate the effects of a {\it small and dense} 
 quenched disorder on the two original Kondo insulators, namely the ferromagnetic 2-leg spin ladder and 
 the S=1 antiferromagnet. We use an Abelian representation of the SU(2) matrices f and g and by
 introducing the usual boson fields ${\Phi}_{\pm}=({\Phi}_{c}\pm{\Phi}_{f})/\sqrt{2}$ for
 the spin sector and their canonical conjugate momenta $\Pi_{\pm}=\partial_x\tilde{\Phi}_{\pm}$, we obtain:
   \begin{eqnarray}
 H^*&=&\sum_{\nu=+,-}\int dx\  \frac{u_{\nu}}{2 K_{\nu}}:{(\partial_x\Phi_{\nu})}^2:+\frac{u_{\nu}K_{\nu}}{2} :{(\Pi_{\nu})}^2:\\ \nonumber      
    &+&\int dx\  \frac{cst}{\alpha^2}[\lambda_4\cos(\sqrt{4\pi}\Phi_-)\\ \nonumber
    &+&2\lambda_5\cos(\sqrt{4\pi}\tilde{\Phi}_-)-\lambda_6\cos(\sqrt{4\pi}{\Phi_+)}]\cos(\sqrt{2\pi}\Phi_c^c)
    \end{eqnarray}
   As seen before, the Kondo term does not change the physics and
   it has not been introduced; then, the parameter $K_{-}$ takes 
   into account the fact that the SU(2) spin symmetry is explicitly 
   broken for a strong ferromagnetic Kondo coupling: in that case $K_{-}$
   is large and the massive excitations are reduced to the doublet $S^z=\pm 1$
   well-described by the term $\cos(\sqrt{4\pi}\phi_+)$\cite{cinqs}. Finally, we
   want to remember that a ``weak $\lambda_k$-coupling'' approach still gives good results
   concerning the S=1 antiferromagnet ground state.
   \\
   Here, we apply renormalization group methods, first used
   by Giamarchi and Schulz\cite{treize} (in the context of randomness in a TL liquid). We introduce the complex random impurity potential,
   \begin{equation}
   H_{imp}=\sum_{\sigma}\int dx\ \xi(x)c^{\dag}_{\sigma L}c_{\sigma R}\qquad +hc
   \end{equation}
   with the Gaussian distribution:
   \begin{equation}
   P_{\xi}=\exp(-D_{\xi}^{-1}\int dx\ {\xi}^{\dag}(x){\xi}(x))
   \end{equation}
   We can omit forward scatterings, because the k=0 random potential just
   renormalizes the chemical potential and it does not affect the fixed point properties. In order to deal with the quenched disorder, we use the well-known {\it replica} trick\cite{quatorze}. Due
   to the definition of ${\xi}(x)$, we are limited to a weak randomness treatment
   and we incorporate only the first contribution (order) in $D_{\xi}$. Then, there
   is no coupling between different replica indices, which will be omitted
   below. By applying a standard renormalization group analysis up to the lowest
   order in $\lambda_k$ and $D_{\xi}$, we obtain the complete flow:
   \begin{eqnarray}
   \label{une}
    \frac{d D_{\xi}}{dl}&=&(3-\frac{K_+}{2}-\frac{K_-}{2}-K_{\rho})D_{\xi}\\ 
    \frac{d\lambda_4}{dl}&=&(2-K_- -\frac{K_{\rho}}{2})\lambda_4\\
    \label{deuxe} 
    \frac{d\lambda_5}{dl}&=&(2-\frac{1}{K_-}-\frac{K_{\rho}}{2})\lambda_5\\ 
    \frac{d\lambda_6}{dl}&=&(2-K_+ -\frac{K_{\rho}}{2})\lambda_6\\ 
    \label{un}
    \frac{d K_+}{dl}&=&-\frac{1}{2}(D_+ +\lambda_6^2)K_+^2\\
    \label{trois} 
    \frac{d K_-}{dl}&=&-\frac{1}{2}(D_- +\lambda_4^2)K_-^2+2\lambda_5^2 K_-^2\\
    \label{deux}
    \frac{d K_{\rho}}{dl}&=&-\frac{1}{2}u_{\rho}[(\frac{D_-}{u_-} + \frac{D_+}{u_+})+(\frac{(\lambda_4^2+4\lambda_5^2)}{u_-}+\frac{\lambda_6^2}{u_+})]K_{\rho}^2
   \end{eqnarray}
   with the notations: $D_{\nu}=2\frac{D_{\xi}\alpha}{\pi u_{\nu}^2}[\frac{u_{\nu}}{u_{\rho}}]^{K_{\rho}}$
   and $\tilde{\lambda}_{\nu}=\frac{\lambda_{\nu}}{2\pi u_{\nu}}$ and where we
   did not display the irrelevant equations to the following discussions. We can
   particularly notice that in this problem, there is no term like $D_{\xi}\lambda_{(4,6)}$
   generated by perturbation. Finally, the fixed point
   properties clearly depend on the initial conditions, on the pure system. 
   \\
   For instance, if we consider the 2-leg spin ladder situation, we initially have $2-K_- -\frac{K_{\rho}}{2}>0$, $2-\frac{1}{K_-}-\frac{K_{\rho}}{2}>0$ 
    and $2-K_+ -\frac{K_{\rho}}{2}>0$. By using the above equations, we can easily conclude that, in this case, $D_{\xi}$
   scales to the strong coupling regime before the couplings $\lambda_{\nu}$: the Haldane
   gap vanishes and $K_{\rho}$
   tends to zero. Although in the strong coupling regime, our treatment for randomness breaks down, we can 
   guess the fixed point properties. Indeed, the charge gap $\Delta_c\propto\lambda_k^2$ 
   is small and it seems plausible to expect that the TL
   liquid will be (still) quenched by a strong disorder, to give 
   a transition into the famous Anderson localization state\cite{aa}. 
   \\
   Now, if we branch a strong ferromagnetic Kondo coupling by decreasing the hopping
   strength $t$, we initially have $2-K_- -\frac{K_{\rho}}{2}<0$, $2-\frac{1}{K_-}-\frac{K_{\rho}}{2}>0$ 
     and $2-K_+ -\frac{K_{\rho}}{2}>0$ and $K_->>1$ since the SU(2) spin
    symmetry is explicitly broken onto $U(1)\times{\cal Z}_2$. By using
    the eqs. (\ref{un}), (\ref{deux}), we can clearly see that 
    $K_+$ and $K_{\rho}$ are also reduced by the randomness and  gaps
    open in $\Phi_+$ and $\Phi_c^c$: finally, $K_{-}\rightarrow +\infty$ at the fixed
    point and the disorder is reduced to zero. The coupling $\lambda_6$ scales to the strong coupling limit and
    we can check that the Haldane gap is then given by the single instantonic
    term $\cos(\sqrt{4\pi}\phi_+)$. Then, the Haldane
    gap is maintained and it is not surprising 
    if we remember that the charge gap 
    $\Delta_c\propto\lambda_k^2$ is large in the pure system and
    finally it is even larger than the Haldane gap ($\Delta_S\propto 1/2.\lambda_k^2$): the electrons
    are quite quenched by magnetic impurities and they are not sensitive to a small quenched
    disorder.
 
   Summarizing, for a half-filled band, the 2-leg spin ladder and the S=1 antiferromagnet occur
   as new original Kondo insulators; both, they yield a short-range 
   RVB ground state and massive singlet-triplet excitations. But, these two similar
   Kondo insulators do not equivalently react by hardly hole-doping
   or by applying a quenched and dense disorder. First, a strong ferromagnetic Kondo coupling is essential to maintain the Haldane
   gap and to prevent a transition into an Anderson localization state. However, we expect the step of a charge localization 
   in the transport properties, whatever the ratio $\lambda_k/t$: localization by the ``randomness'' in the spin-ladder context and localization by
   the spins in the case of a strong ferromagnetic Kondo coupling. Second, by hardly hole-doping, the ground state becomes very dependent on the 
   parameter $\left|\lambda_k\right|/t$. Indeed, for a ratio
  $\left|\lambda_k\right|/t<<1$, the antiferromagnetic correlations are enhanced
 due to the emergence of an incommensurate RKKY interaction. Conversely, when 
 the ferromagnetic Kondo coupling is large or $\left|\lambda_k\right|/t>1$, we 
 expect another curious change by hardly hole-doping: the residual spin system, prevalently formed by S=1 spins
 yields a ferromagnetic ground state: it minimizes the kinetic energy. Finally, if we 
 remove a $S_f$-spin of the lattice instead of an electron, we predict the 
 same conclusions concerning the spin sector, since $\lambda_3=0$ in that case too, but the
 system remains insulator because the spins can not move.

   \end{document}